\documentclass[reprint,aps,prd,twocolumn,notitlepage,nofootinbib,
showpacs,preprintnumbers,superscriptaddress]{revtex4-1}
\usepackage[T1]{fontenc}
\usepackage[utf8]{inputenc}
\usepackage{bm}
\usepackage{amsmath}
\usepackage{amssymb}
\usepackage{esint}
\usepackage{color}
\usepackage{graphicx}% Include figure files
\PassOptionsToPackage{normalem}{ulem}
\usepackage{ulem}

\usepackage{times}
\usepackage{hyperref}
\usepackage{amsthm,bm,mathrsfs,stmaryrd,amsmath}

\newcommand{\be}{\begin{equation}}
\newcommand{\ee}{\end{equation}}
\newcommand{\ba}{\begin{eqnarray}}
\newcommand{\ea}{\end{eqnarray}}

\def\pa{\partial}
\def\D{\Delta}

\def\be{\begin{eqnarray}}
\def\ee{\end{eqnarray}}
\def\D{\Delta}

\def\D{\Delta}

\def\pa{\partial}

\newcommand\<\langle
\renewcommand\>\rangle

\begin{document}

\title{Null bootstrap for non-Hermitian Hamiltonians}

\author{Wenliang Li}
\email{liwliang3@mail.sysu.edu.cn}
\affiliation{School of Physics, Sun Yat-sen University, Guangzhou 510275, China}

\begin{abstract}
A stable physical system has an energy spectrum that is bounded from below. 
For quantum systems, the dangerous states of unboundedly low energies should decouple and become null. 
We propose the principle of nullness and apply it to the bootstrap study of 
Hermitian and non-Hermitian anharmonic oscillators. 
\end{abstract}

\maketitle 
\tableofcontents
\section{Introduction}
The bootstrap is an ambitious program that attempts to classify and solve physics 
by basic principles and consistency relations of observables. 
The governing role of basic principles can be viewed as a reduction at the conceptual level, 
instead of the usual model-building level. 
On the other hand, the emphasis on the consistency of all observables, 
rather than a construction based on some smallest building blocks, 
is closer to the spirit of emergence. 

The first concrete bootstrap program came from early investigations of the strong nuclear force. 
In the 1960's, Chew proposed the S-matrix bootstrap to determine the scattering amplitudes 
by unitarity  and some subtle analyticity conditions. 
This is not a simple program due to the strong coupling nature  
and became dormant later. 
In the 1970's, the bootstrap idea also appeared in the study of conformal field theory (CFT) \cite{Ferrara:1973yt,Polyakov:1974gs}. 
In two dimensions, the conformal bootstrap program has achieved considerable progress \cite{Belavin:1984vu,DiFrancesco:1997nk}. 
One of the most beautiful results is the complete classification of 2d minimal-model CFTs. 
They describe the critical behaviour of statistical physics models, such as the Ising, Yang-Lee and Potts models. 

The conformal bootstrap in $d>2$ is more challenging and thus not many results had been obtained   
until the modern numerical approach was proposed in \cite{Rattazzi:2008pe}. 
In this seminal work, the unitarity assumption and the crossing equation are formulated as inequalities, 
so one can rule out the inconsistent theory space. 
%Unitarity means that the norm of a state is either zero or positive. 
One of the impressive results is the most precise determination of the 3d Ising critical exponents 
\cite{ElShowk:2012ht, El-Showk:2014dwa, Kos:2014bka, Kos:2016ysd}. 
More recently, the positivity principle 
\footnote{By positivity constraints, we mean some quantities should be either positive or zero.}  
has also been applied to the studies of matrix models and quantum mechanical systems 
\cite{Anderson:2016rcw,Lin:2020mme,Han:2020bkb,Han:2020,Kazakov:2021lel,Berenstein:2021dyf,Bhattacharya:2021btd,Aikawa:2021eai,Berenstein:2021loy,Tchoumakov:2021mnh,Aikawa:2021qbl,Du:2021hfw,Lawrence:2021msm,Bai:2022yfv,Nakayama:2022ahr}. 
%Below we will refer these studies as the positive bootstrap. 

In the 2d bootstrap studies, 
positivity does not play such a dominant role. 
In particular, since statistical physics models do not need to obey reflection positivity, 
they can be naturally related to non-unitary quantum systems. 
A notable example is the 2d bootstrap results of 
the Yang-Lee edge singularity \cite{Cardy:1985yy,Cardy:1989fw}. 
To study the non-positive systems in $d>2$, we are led to the following question: 

{\it 
Is there a bootstrap principle for non-positive systems?}

\section{The principle of nullness}
Let us start with the two-dimensional conformal field theories. 
They are among the most well-understood nonperturbative quantum systems at strong coupling.  
A particularly simple class of the 2d CFTs is the minimal models $\mathcal M(p,p^\prime)$ \cite{DiFrancesco:1997nk}.  
Furthermore, the discrete series of unitary CFTs corresponds to 
the minimal models with $|p-p^\prime|=1$ \cite{Friedan:1983xq}. 
In unitary theories, the norm of a state is either positive or zero. 
Positivity can be viewed as a basic principle for 
bootstrapping these unitary systems. 

Why does the unitarity condition lead to a subset of minimal models? 
Since the minimal models are characterized by the existence of many null states, our interpretation is: 
The would-be dangerous states are removed as  
they become null and orthogonal to the positive norm states. 
In this way, the resulting truncated space of states is compatible with the unitarity condition.

However, unitarity is not a necessary property of physical systems. 
A celebrated example is the non-unitary $i\phi^3$ theory describing the critical behaviour of the Yang-Lee edge singularity 
\cite{Fisher:1978pf}. 
In 2d, the corresponding CFT is the minimal model $\mathcal M(5,2)$. 
There are many more examples of non-unitary physical systems, such as polymer and percolation physics. 
The non-unitary CFTs can be associated with statistical physics models that are not reflection positive. 
More generally, the positivity condition can be violated in analytically continued or complex systems. 

Is there a principle for these equally important non-positive systems? 
The 2d minimal models clearly suggest a principle based on nullness, i.e. 
there exist many null states. 
\footnote{Some properties of non-minimial models are also constrained by the null state conditions. }
In 2d, the principle of nullness is more general than the principle of positivity
because the unitary models are a subset of minimal models 
when the central charge is less than one. 
In 3d, the phenomena of operator decoupling have also been noticed 
in the positive bootstrap studies of the Ising CFT \cite{ElShowk:2012ht, El-Showk:2014dwa}. 

As concrete examples, we will consider some interacting quantum mechanical (QM) systems. 
As mentioned in the introduction, 
several Hermitian systems have been recently revisited using the positivity principle. 
In this work, we will use the null bootstrap to solve the anharmonic oscillator with a quartic term. 
To show its broader applicability, we will also consider a non-Hermitian Hamiltonian, 
which is closely related to the Yang-Lee edge singularity mentioned above. 

For the harmonic oscillator, 
the null state conditions can be derived from a well-known fact:  
The vacuum state is annihilated by lowering operators. 
Should the null states present in more complicated problems? 
In an algebraic formulation of quantum mechanics, 
one can construct lowering operators 
that generate lower energy eigenstates from a given eigenstate. 
If the energy spectrum is unbounded from below due to these lowering operators, 
then the stability of the physical system is destroyed. 
\footnote{In a more general context, the boundedness requirement on the real part of energies 
can be associated to different kinds of spectra, such as twists. }
In addition, one can extract an infinite amount of energies by repeatedly lowering the energy of the quantum system. 
Therefore, we need a termination mechanism for a physical space of quantum states. 
\footnote{The state space is not necessarily a Hilbert space 
because the norm of a state is allowed to be negative. }
This can be naturally realized by null state conditions. 
If the lowering operators are associated with Lie algebra generators, 
the null state conditions reduce to the highest weight conditions in representation theory.

Roughly speaking, there are two types of null state conditions: 
\begin{enumerate}
\item
The first type is related to the model definitions, 
such as $\mathcal O(H-E)|\psi_E\>=0$ from the Schr\"{o}dinger equation
and ``equations of motion'' in field theory. 
\item
The second type removes the dangerous states in the spectra. 
\end{enumerate}
In this work, the first type will be called trivial as they follow directly from the definition of Hamiltonian.
\footnote{Alternatively, they may be viewed as operator equations or properties of a concrete representation of an abstract operator algebra. 
Note that the first type is non-trivial in the conformal bootstrap. }
An unstable model may have unbounded spectra in all choices of state spaces,  
so the null state conditions of the first type also remove dangerous states 
and the division is not completely definite.
The two types of null state conditions are unified into the principle of nullness. 

\section{The null bootstrap for quantum mechanics}
\label{sec-null-bootstrap-QM}
In the bootstrap approach, 
we study the relationships of physical observables
and extract the predictions from some basic principles. 
For a QM system, we will assume that the Hamiltonian is diagonalizable 
and consider the expectation values associated with 
an energy eigenstate $|\psi_E\rangle$. 
By definition, they satisfy the consistency relations
\be
\langle\psi_E|\mathcal OH|\psi_E\rangle
=E\langle\psi_E|\mathcal O|\psi_E\rangle
=\langle\psi_E| H \mathcal O|\psi_E\rangle\,,
\label{H-relation}
\ee
where the Hamiltonian will be assumed to take the form $H=p^2+V(x)$. 
We assume that the position and momentum operators satisfy the canonical commutation relation
$[x,p]=i\hbar$, where $\hbar=1$. 
The choice of a state corresponds to a representation of the operator algebra. 
The space of states is generated by the action of operators on $|\psi_E\>$. 
Below we will also use a shorthand notation 
for the expectation values associated with the energy eigenstate $\psi_E$
\be
\langle\mathcal O\rangle_E=\langle\psi_E|\mathcal O|\psi_E\rangle\,.
\ee 
The first equality in \eqref{H-relation} can be easily derived from the Schr\"{o}dinger equation
$H|\psi_E\rangle=E|\psi_E\rangle$. 
To obtain the second equality of \eqref{H-relation}, we need to specify an inner product. 
\footnote{The notion of null states should be independent of the choice of inner products, 
but it is useful to introduce inner products to detect them. }
A natural choice will depend on the properties of the Hamiltonian $H$. 
If $H$ is Hermitian, 
we will use the standard Hermitian inner product, i.e.  
$\langle\psi_1|\psi_2\rangle^{\mathcal {H}}=\int \mathrm{d}x\, [\psi_1(x)]^\ast\,\psi_2(x)$ in the position representation. 
For a non-Hermitian Hamiltonian 
with space–time reflection and matrix transposition symmetries, 
we will instead use the $\mathcal P\mathcal T$ inner product, i.e.    
$\langle\psi_1|\psi_2\rangle^{\mathcal P\mathcal T}=C\int \mathrm{d}x\, [\psi_1(-x)]^\ast\psi_2(x)$, 
where $C$ is a constant. 
In this way, we also have \eqref{H-relation} and the orthogonality of eigenstates is preserved in the non-Hermitian case. 

To make physical predictions using consistency relations, 
the bootstrap approach will follow some basic principles. 
For Hermitian systems, one can use the unitarity condition and the associated positivity constraints. 
For non-Hermitian systems, the norm of a state is allowed to be negative, 
then the principle of positivity can be violated. 
As proposed above, a more general principle is to impose the existence of many null states. 
They satisfy
\be
\langle\mathcal \psi_\text{test}| \psi_\text{null}\rangle
=\langle\mathcal O_\text{test}\,\mathcal O_\text{null}\rangle_E
=0\,,
\label{null}
\ee
where the test and null states are generated by  
the action of operators $\mathcal O_\text{test}, \mathcal O_\text{null}$ 
on the energy eigenstate $\psi_E$. 
For a given null state $| \psi_\text{null}\rangle=\mathcal O_\text{null}| \psi_E\rangle$, 
the inner product \eqref{null} should vanish for arbitrary test operator $\mathcal O_\text{test}$. 
\footnote{We neglect some minor convention issues associated with inner products 
and  define $\<\psi_\text{test}|=\<\psi_E|\mathcal O_\text{test}$. } 
We will not consider the dual version $\langle\psi_\text{null}|\mathcal \psi_\text{test}\rangle=0$, as it is not independent. 

To show that the principle of nullness captures both positive and non-positive physics, 
we will apply it to some classic examples of Hermitian and non-Hermitian quantum systems in Sec. \ref{application}. 
Below we discuss the procedure of the null bootstrap in more detail.   

\subsection{Complete vs reduced procedure}
In the complete procedure of the null bootstrap, 
one first determines the spectrum  
and then imposes the null state conditions on the dangerous eigenstates.  
In terms of expectation values, 
the spectrum $\{E_k\}$ is defined by the consistent rewriting of \eqref{H-relation} into  
the Schr\"{o}dinger-like form 
\be
\langle\mathcal \psi_\text{test}|(H-E_k)|\psi_k\rangle=
\langle\mathcal O_\text{test}(H-E_k)L_k\rangle_E
=0\,,
\label{Schrodinger}
\ee 
where the eigenstates are labelled by the abstract index $k$. 
Note that $\mathcal O_\text{test}$ is again arbitrary and  
the ladder operator $L_k$ connects different eigenstates, 
i.e. $L_k|\psi_E\>=|\psi_k\>$. 
Then we impose the null state conditions on the dangerous states 
with unboundedly low energies. 

As an explicit example, let us consider the simple harmonic oscillator $H=p^2+x^2$. 
\footnote{The discussion here was partly inspired by the positive bootstrap in \cite{Aikawa:2021qbl}.}
The rewriting of \eqref{H-relation} into \eqref{Schrodinger} 
leads to the spectrum $E_k=E+2k$ where $k$ is an integer. 
The lowering operators can be represented by $L_{-n}=(x+ip)^{n}$ with $n=1,2,\cdots$. 
\footnote{The lowering operators are not completely determined by the Schr\"{o}dinger-like equation \eqref{Schrodinger} 
because one can always add a trivial null state $\mathcal O(H-E)|\psi_E\>$ . }
Since a null state has zero norm and 
$\< (x-ip)^n(x+ip)^n\>_E
=\< 1\>_E \prod_{k=0}^{n-1}(E-2k-1)$,   
the null state conditions on the dangerous states with unboundedly low energies 
imply $E_n=2n+1$ with $n=0,1,2,\cdots$. 
In this way, we solve the exact energy spectrum of the harmonic oscillator 
using the complete procedure of the null bootstrap. 

%\subsection{Reduced procedure}
We can also consider a reduced procedure. 
In general, the lowering operators are expected to be much more complicated than 
polynomials of finite degrees,  
so it is hard to determine the precise spectrum in the first step of the complete procedure. 
Nevertheless, it is possible to impose the existence of many null states 
without specifying the explicit expressions of the dangerous states. 
{\it A crucial point is that certain superposition of many null states can be well approximated by simple polynomials. }
Therefore, we can carry out the search for null states using a simple ansatz 
in the $\eta$ minimization described below.  

\subsection{$\eta$ minimization}
In the numerical implementation, we will consider truncated null state conditions
\be
\langle \psi_\text{test}^{(L)}| \psi_\text{null}^{(K)}\rangle\approx0\,, 
\label{truncated-null}
\ee
where the candidate null states are truncated to order $K$ 
and the test states $\langle\psi_\text{test}|$ are truncated to order $L$
\be
| \psi_\text{null}^{(K)}\rangle=
\sum_{m=0}^{K}\sum_{n=0}^{K-m} a_{m n }\,x^{m}(ip)^{n}
|\psi_E\rangle\,,
\label{null-K}
\ee
\be
\langle\psi_\text{test}^{(L)}|=\sum_{m=0}^{L}\sum_{n=0}^{L-m} b_{m n }\,\langle\psi_E|x^{m}(ip)^{n}\,.
\label{test-L}
\ee
Here we assume that the Hamiltonian is expressed in terms of the position and momentum operators $x, p$, 
but it is straightforward to generalize \eqref{null-K}, \eqref{test-L} to the case with more operators. 
The orders $K, L$ are associated with the polynomials in $x, p$ acting on the energy eigenstate $\psi_E$.  

To have a nontrivial system of constraints, the number of independent constraints 
should be greater than that of independent parameters. 
Usually we are not able to solve all truncated null state conditions 
at the same time, so we write $\approx$ in \eqref{truncated-null}. 
Now we introduce the $\eta$ function
\be
\eta=\sqrt{\sum_{m=0}^{L}\sum_{n=0}^{L-m}\left|\frac{1}{m!n!}\frac{\pa\langle \psi_\text{test}^{(L)}|\psi_\text{null}^{(K)}\rangle}{\pa{b_{mn}}}\right|^2}\,,
\label{eta-fun}
\ee
which measures the violation of null constraints by 
summing the squares of different components.  
The null constraints of higher orders are suppressed by $(m!n!)^{-2}$.
\footnote{One may consider more sophisticated weights. }
To avoid the trivial solution $a_{mn}=0$, we will impose the constraint $\sum_{m,n}a_{mn}=1$. 
For a given set of independent expectation values, 
the $\eta$ minimization 
can be carried out easily by linear least squares methods. 
\footnote{In the conformal bootstrap, 
the $\eta$ minimization was proposed in \cite{Li:2017ukc} as a more systematic formulation of the truncation approach initiated in \cite{Gliozzi:2013ysa}. 
The truncation of operator product expansions can be reinterpreted as approximate null operators. 
As shown in \cite{Li:2021uki}, one can also rule out the inconsistent solutions using the $\eta$ minimization. 
Recently, the $\eta$ minimization was implemented with 
reinforcement-learning algorithms \cite{Kantor:2021kbx,Kantor:2021jpz}. 
}

As a simple example, 
let us consider again the harmonic oscillator $H=p^2+x^2$. 
After solving \eqref{H-relation}, all the expectation values $\<x^mp^n\>_E$ are functions of $E$. 
If we choose $K=1$ and $L\geq1$, the minimization of the $\eta$ function has a unique zero at $E=1$ 
and the null state is given by 
$|\psi_\text{null}^{(1)}\>=(x+ip)|\psi_E\>$. 
For $K=2$ and $L\geq 2$, we find two zeros of $\eta_\text{min}$ and they are at $E=1, 3$. 
One can see that the $\eta$ minimization leads to exact results in this simple example. 
In addition, a larger $K$ captures more excited states. 
For $K=2$, we impose an additional constraint $a_{00}=0$ to remove the trivial solution 
$\mathcal O^\text{trivial}_\text{null}=H-E$. 
\footnote{We have imposed the constraint $\sum_{m,n}a_{mn}=1$. 
In a different weighted sum, one may also find solutions with $E=-1,-2,-3,\cdots$, 
which correspond to a spectrum bounded from above. 
In the dual case, the null conditions are for the positive energy eigenstates. 
From the perspective of wave functions, the boundary conditions are imposed at $x\rightarrow \pm i \infty$. 
It would be fascinating to further study the relations between the choices of coordinates and null state conditions. } 

In general, an energy eigenstate can correspond to a nontrivial lowering operator, 
so it is much more complicated than finite degree polynomials. 
Can we still study the null state conditions using the polynomial ansatz? 
The crucial point mentioned earlier is that the null states form a high-dimensional subspace for physical solutions. 
\footnote{The null operators form a left ideal in the operator algebra because the action of any operator on a null state also gives a null state. From the mathematical perspective, the null bootstrap can be viewed as a classification program based on the ideals in operator algebra.}
In the $\eta$ minimization, we are allowed to consider a linear combination of all of them  
and the polynomial ansatz becomes a good approximation. 
The minimized $\eta$ function is particularly small around the physical solutions, 
so we can extract the precise predictions using the reduced procedure of the null bootstrap. 

Below we apply the simple $\eta$ minimization to Hermitian and non-Hermitian anharmonic oscillators. 
We obtain high precision results despite severe truncations. 

\section{Application to anharmonic oscillators}
\label{application}
In this section, we will study the Hamiltonian 
\be
H=p^2+V(x)
\ee
using the null bootstrap. 
We will consider two different potentials $V(x)$. 
\subsection{$V(x)=x^2+g x^4$}
Let us add a quartic term to the harmonic oscillator. 
The Hamiltonian becomes 
\be
H=p^2+x^2+g x^4\,,
\label{H4}
\ee
where we will set $g=1$ in the concrete computation. 
This simple model can be viewed as the $0+1$ dimensional version of the $\phi^4$ quantum field theory 
and has long served as a testing problem for new field theory methods \cite{Bender:1969si}.

Using the consistency relations \eqref{H-relation}, we can express $\<x^m p^n\>_E$ in terms of 
$\{E, \<x^2\>_E, \<1\>_E\}$,
where $m,n=0,1,2,\cdots$. 
\footnote{For parity symmetric solutions, we have $\<x^m p^n\>_E=0$ when $m+n$ is an odd number. 
It would also be interesting to consider the analytic continuation of $m,n$ to the non-integer or complex domain.} 
The expectation value $\<1\>_E$ determines the normalization of the inner product 
and we will set $\<1\>_E=1$ for convenience. 
The null bootstrap is not sensitive to this normalization convention as long as $\<1\>_E\neq 0$. 
For instance, the results remain essentially the same even if we set $\<1\>_E=-1$. 

In the null bootstrap, we minimize the $\eta$ function with $K=1,2,3,4$ and $L=K+2$. 
\footnote{In this example, the system is under-constrained if $L=K+1$. }
To remind the reader, the candidate null operator $\mathcal O_\text{null}$ is approximated by a degree-$K$ polynomial in $x,p$. The degree of the test operator $\mathcal O_\text{test}$ is given by $L$. 
As long as the system is over-constrained, 
we can remove some high order constraints and the results will be more precise, 
but we do not present them for the sake of simplicity. 
We can also consider more constraints by choosing a larger $L$. 
The results are qualitatively the same if $L$ is not significantly greater than $K$. 
In this way, we verify that the local minima are not coincidences 
and the null subspace is indeed of high dimensions.  

\begin{figure}[h!]
\begin{center}
\includegraphics[width=8cm]{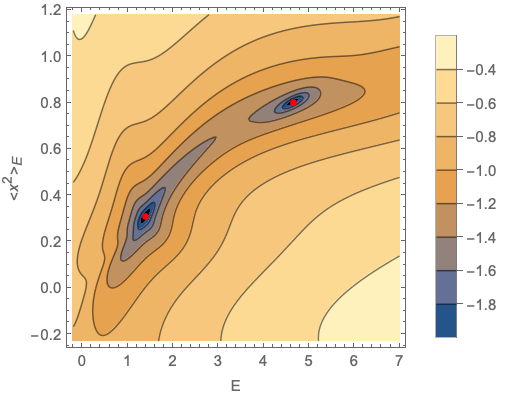}
\caption{The $\eta_\text{min}$ landscape for the Hermitian Hamiltonian $H=p^2+x^2+x^4$ with $K=2, L=4$.  
%In \eqref{truncated-null}, 
%the null states are approximated by degree-2 polynomials in $x,p$ and 
%the test state is generated by degree-4 polynomials in $x,p$. 
The two local minima correspond to the ground state and the first excited state. 
We plot $\log_{10} \eta_\text{min}$ as a function of the energy $E$ and the expectation value $\<x^2\>_E$. 
As a height function, $\eta_\text{min}$ increases rapidly outside of the presented region. 
We use black color to denote the region with $\eta_\text{min}<10^{-2}$. 
The exact solutions are indicated by the red points. } 
\label{figure-K=2}
\end{center}
\end{figure}

\begin{figure}[h!]
\begin{center}
\includegraphics[width=8cm]{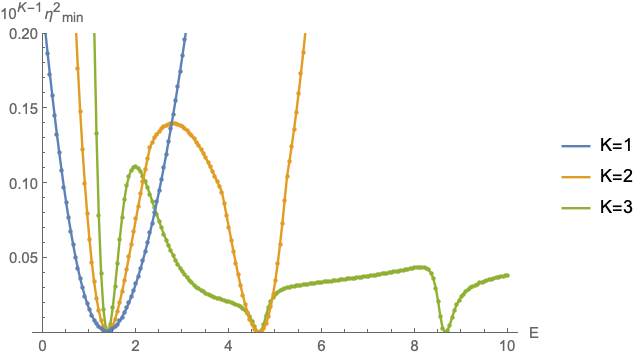}
\caption{The minimized $\eta$ as a function of energy $E$ 
with the Hermitian Hamiltonian $H=p^2+x^2+x^4$ and $L=K+2$.  
The degrees of the polynomial approximations for null states are $K=1,2,3$. 
We carry out the minimization with respect to $\langle x^2\rangle$ as well. 
As we increase $K$, 
the minima give sharper predictions and capture more excited states. 
Here we change the weight in \eqref{eta-fun} to one.  
} 
\label{figure-different-K}
\end{center}
\end{figure}

In Fig. \ref{figure-K=2}, we present the $K=2$ results of $\log_{10} \eta_\text{min}$ 
as a function of the two independent parameters, i.e. $E$ and $\<x^2\>_E$. 
The two local minima correspond to the ground state and the first excited state. 
For a given $E$, the $\eta$ minimization with respect to $\<x^2\>_E$ is quite straightforward, 
so we present the $\eta_\text{min}$ at different $E$ with $K=1,2,3$ in Fig. \ref{figure-different-K}. 
As we increase $K$, more exited states are captured by the $\eta$ minimization. 
We further summarize the numerical predictions in Table \ref{tableEn} and \ref{tablex2}. 
One can see that the errors in the $E$, $\<x^2\>_E$ results decrease rapidly with $K$. 
Here the reference results are derived from the diagonalization of the Hamiltonian 
in a truncated basis of $60$ harmonic oscillator eigenstates. 
As in the case of the harmonic oscillator, 
we set $a_{00}=0$  at $K=4$ in order to avoid the trivial solution $\mathcal O^\text{trivial}_\text{null}=H-E$.

\begin{table}[h]
\begin{tabular}{|c|c c c c|}
\hline
$\Delta E_n^{(K)}$ & $K=1$ & $K=2$ & $K=3$  &$K=4$ \\
\hline
$n=0$ &$-1\times10^{-3}$& $-2\times10^{-3}$ & $-4\times10^{-10}$ & $-7\times10^{-12}$\\
$n=1$ &&$3\times10^{-3}$ &$-3\times10^{-5}$&$2\times10^{-11}$ \\
$n=2$ &&&$5\times10^{-6}$&$6\times10^{-7}$ \\
$n=3$ &&&&$1\times10^{-7}$ \\
\hline
\end{tabular}
\caption{The null bootstrap predictions 
for the low-lying energies of anharmonic oscillator with the Hermitian potential $V(x)=x^2+x^4$ and $L=K+2$. 
We present the first digit of the error $\D E^{(K)}_n=E^{(K)}_n-E^\star_n$.  
The reference energies  $E_0^\star=1.39235164153029$, $E_1^\star=4.648812704212$,  
$E_2^\star=8.65504995776$ and
$E_3^\star=13.1568038980$  
are computed from 
diagonalizing the Hamiltonian with size $60\times 60$ 
in the basis of harmonic oscillator eigenfunctions. }
\label{tableEn}
\end{table}

\begin{table}[h]
\begin{tabular}{|c|c c c c|}
\hline
$\Delta \langle x^2\rangle_n^{(K)}$ & $K=1$ & $K=2$ & $K=3$  &$K=4$ \\
\hline
$n=0$ &$-1\times10^{-2}$& $-1\times10^{-4}$ & $2\times10^{-9}$ & $1\times10^{-11}$\\
$n=1$ &&$1\times10^{-3}$ &$-1\times10^{-6}$&$1\times10^{-11}$ \\
$n=2$ &&&$-3\times10^{-6}$&$6\times10^{-8}$ \\
$n=3$ &&&&$2\times10^{-8}$ \\
\hline
\end{tabular}
\caption{The null bootstrap predictions 
for $\langle x^2\rangle_n$ with the Hermitian potential $V(x)=x^2+x^4$ and $L=K+2$. 
We present the first digit of the error $\Delta \langle x^2\rangle_n^{(K)}= \langle x^2\rangle^{(K)}_n- \langle x^2\rangle^\star_n$.  
The reference values  
$\langle x^2\rangle_0^\star=0.3058136507176$, 
$ \langle x^2\rangle_1^\star=0.8012505955411$,  
$ \langle x^2\rangle_2^\star=1.155440519200$ and
$ \langle x^2\rangle_3^\star=1.4675232154$  
are computed using approximate wave functions from the diagonalization of the Hamiltonian with size $60\times 60$ 
in the harmonic oscillator basis. }
\label{tablex2}
\end{table}
 
Instead of rigorous bounds in the positive bootstrap, 
the null bootstrap gives an $\eta_\text{min}$ landscape. 
At the price of rigorousness, 
the null bootstrap results are more precise than 
those of the positive bootstrap.  
For instance, the size of the $E_0$ error bar from the rigorous island is
$\delta E_0^{(K)}=\infty, 3\times 10^{-2}, 1\times 10^{-4}, 5\times 10^{-10}$ 
with $K=1,2,3,4$. 
\footnote{Recently, the quartic anharmonic oscillator has been revisited in the positive bootstrap \cite{Han:2020bkb,Bhattacharya:2021btd} in the basis of $\{x^n\}$ and \cite{Du:2021hfw,Aikawa:2021qbl} in the basis of $\{x^mp^n\}$. 
In \cite{Du:2021hfw}, the corresponding results are more precise  
because they define $K$ as $\max\{m,n\}$. 
The degrees of their polynomials are twice of ours. }
In the positive bootstrap, one asks if the positivity condition 
\footnote{To be more precise, some matrices should be positive semidefinite as null states are allowed. } 
is violated and there is no smooth change in the answer. 
\footnote{But one might introduce a measure of positivity violation based on the dangerous states, 
such as the lowest eigenvalues of the should-be-positive matrices. 
See also the navigator function in the conformal bootstrap \cite{Reehorst:2021ykw}.  }
Then it requires some care to determine the boundaries of the positive regions, 
especially when an isolated region has a tiny size. 
In the null bootstrap, it is relatively easy to find the local minima 
thanks to the smooth $\eta_\text{min}$ landscape. 
\footnote{We also consider a negative coefficient for the quadratic term. i.e. 
the double-well potential $V(x)=-5x^2+x^4$ with $L=K+2$, 
which was also revisited in the positive bootstrap \cite{Bhattacharya:2021btd,Berenstein:2021loy,Nancarrow:2022wdr}. 
The $K=1$ truncation does not give a physical minimum due to the near-degeneracy of the first two energies 
with $E_0\approx-3.4101428$, $E_1\approx -3.2506754$. 
For $K=2,3$, we find $K$ local minima, corresponding to the first $K$ energy levels. 
For $K=4$, we obtain 6 minima associated with the low energy eigenstates. 
The errors are $\D E_{n}=\{-1\times 10^{-4}, 3\times 10^{-6}, -1\times 10^{-3}, 3\times 10^{-4}, 3\times 10^{-1}, 1\times10^0\}$. 
Using the results of $E$ and $\langle x^2\rangle_E$, 
the ladder operators and matrix elements can be determined as in the case of \eqref{H4}. 
The precision of the double well predictions is lower than that of \eqref{H4} with $g=1$, 
but still unexpectedly good, especially that of $E_1$. }

\begin{figure}[h!]
\begin{center}
\includegraphics[width=8cm]{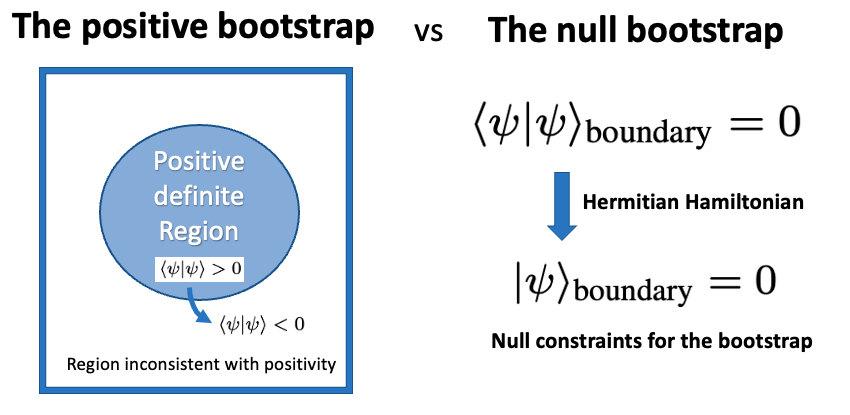}
\caption{The relation between the positive and null bootstrap methods. 
For Hermitian Hamiltonians, the positive bootstrap results can be deduced by the null bootstrap method. 
Along a path in the parameter space, the norm of some state $|\psi\rangle$ changes sign 
after crossing the boundary of the positive region. 
The zero-norm states on the boundary should be null in the Hermitian theories, 
so the positive boundary can be studied directly by the null bootstrap. } 
\label{positive-null}
\end{center}
\end{figure}

%We have shown that the null bootstrap determines the low-lying properties of 
%the Hermitian anharmonic oscillator. 
%The precision of the results also increases rapidly with the cutoff $K$. 
Before moving to the non-Hermitian case, 
let us explain in Fig. \ref{positive-null} from a different perspective why the positive bootstrap results can be 
derived from the null bootstrap. 
In the positive bootstrap, one finds some regions satisfying 
the positive semidefinite condition.  
For some state, the norm will change sign after crossing a boundary,  
so it has zero norm precisely at the crossing point. 
%As we increase the cutoff parameters, 
%the boundary of an allowed island shrinks to a point, 
%so there are many zero-norm states. 
In Hermitian systems, 
zero-norm states must be null, i.e. they should be orthogonal to the positive norm states. 
\footnote{One can see that by inserting a complete set of basis. 
%\be
%\<\psi|
%|\psi\>_\text{boundary}=0 \Rightarrow |\psi\>=0
%\ee
From $\<\psi|\psi\>=\sum_n\<\psi|\phi_n\>\<\phi_n|\psi\>
=\sum_n |\<\phi_n|\psi\>|^2=0$, we have $\<\phi_n|\psi\>=0$. 
One can further rule out the boundary if the null conditions are not satisfied. }
Therefore, we can directly search for the boundary of positive solutions using the null bootstrap. 
\footnote{In fact, the null-state conditions are stronger than the zero-norm conditions.}
For the case of \eqref{H4}, we expect the positive islands shrink to points as $K$ goes to infinity, 
so they are captured by the local minima of the $\eta$ function. 
\footnote{At finite $K$, the space of states is truncated. 
The boundary of a tiny island is associated with a family of zero-norm states. 
Although an exact solution in a positive island has no zero-norm states in the truncated space, 
the smallness of the island size indicates that the null and the truncated subspaces of states have a short distance, 
measured by the $\eta$ function \eqref{eta-fun}. 
}

\subsection{$V(x)=i g x^3$}
Since we propose that nullness is a more general principle than positivity, 
we should apply the null bootstrap to a non-positive system. 
Here we consider a simple non-Hermitian Hamiltonian
\be
H=p^2+ i g x^3\,,
\label{Hx3}
\ee
which is $\mathcal {PT}$ symmetric, i.e. invariant under the space–time reflection. 
We will set $g=1$ in the concrete computation. 

In the textbooks of quantum mechanics, 
a standard axiom is that a fundamental Hamiltonian is Hermitian. 
\footnote{Note that non-Hermitian Hamiltonians have been widely used in phenomenological descriptions.}  
The Hermiticity assumption ensures that the energy spectrum is real and the Hermitian norm is non-negative. 
Furthermore, the eigenfunctions are guaranteed to be orthogonal and complete. 

For non-Hermitian Hamiltonians, the Hermitian norm does not need to have a definite sign  
\footnote{It is possible to construct a $\mathcal{CPT}$ norm that is positive \cite{Bender:2002vv}, 
but the $\mathcal C$ operator takes a non-trivial form, 
which makes the consistency conditions too complicated. 
Another approach is to consider the V norm (see \cite{Khan:2022uyz} and references therein). },
but Bessis and Zinn-Justin noticed that the Hamiltontian \eqref{Hx3} may have a real energy spectrum.  
Their unpublished numerical study was motivated by the properties of the Yang-Lee edge singularities, 
whose critical behaviour is described by the $i\phi^3$ quantum field theory. 
It was later shown by Bender and Boettcher that the energy spectrum is real for a large class of 
non-Hermitian Hamiltonians with unbroken $\mathcal {PT}$ symmetry \cite{Bender:1998ke}. 
Curiously, an imaginary cubic coupling constant also appeared in the study of Reggeon field theory 
\cite{Gribov:1967vfb}. 

%\footnote{For a $\mathcal {PT}$-symmetric eigenfunction of \eqref{Hx3}, 
%the real and imaginary parts of an eigenfunction is coupled to each other by the $ix^3$ term.  
%The corresponding Schr\"{o}dinger equation can be expressed 
%as a fourth order differential equation for a parity-symmetric function. }

When $g$ is small, 
the Hermitian case \eqref{H4} can be viewed as a deformation of the harmonic oscillator, 
so it seems natural to find qualitatively similar results. 
\footnote{The main difference is that the harmonic oscillator case has only one nontrivial parameter, i.e. $E$, 
but the quartic case \eqref{H4} has two, i.e. $E$ and $\<x^2\>_E$. 
If we require the $g\rightarrow 0$ limit of $\<x^mp^n\>_E$ is regular, 
we obtain constraints on the small $g$ expansion of $E$, $\<x^2\>_E$, 
so the number of independent parameters is reduced. 
}
By contrast, the non-Hermitian case \eqref{Hx3} has no mass term
and cannot be viewed as a deformed harmonic oscillator. 
\footnote{They are connected by analytic continuation in the power of $ix$. } 
Without knowing much about the lowering operators and spectra, 
it is a leap of faith to apply the null bootstrap to the non-Hermitian system \eqref{Hx3}. 
Surprisingly, the results of the $\eta$ minimization again exhibit a similar pattern. 
\footnote{This suggests the existence of ladder operators 
but with the form $[H, a_{\pm}]=f_{\pm}(H)\, a_{\pm}$, 
where $f_{+}(H)$ and $f_{-}(H)$ are functions of the Hamiltonian, so the energy spacing is not constant. 
Furthermore, one may generate an operator algebra by $a_{\pm}$ with $[a_-,a_+]=\hbar$ and define the Hamiltonian $H$ in terms of $a_{\pm}$.}

\begin{table}[h]
\begin{tabular}{|c|c c c|}
\hline
$\Delta E_n^{(K)}$ & $K=1$ & $K=2$ & $K=3$  \\
\hline
$n=0$ &$4\times10^{-4}$& $-8\times10^{-7}$ & $1\times10^{-11}$  \\
$n=1$ &&$2\times10^{-3}$ &$-3\times10^{-9}$ \\
$n=2$ && &$-1\times10^{-4}$ \\
\hline
\end{tabular}
\caption{The null bootstrap predictions 
for the low-lying energies of anharmonic oscillator with the non-Hermitian Hamiltonian $H=p^2+ix^3$ and $L=K+1$. 
The errors are defined as $\D E^{(K)}_n=E^{(K)}_n-E^\star_n$ and we present only the first digits.  
The reference energies $E_0^\star=1.156267072$, 
$E_1^\star=4.109228752$, $E_2^\star=7.562 273 854$  
were obtained with the Runge-Kutta method in \cite{Bender:2007nj}. 
A more precise value $E_0^\star=1.156267071988$ is computed 
by diagonalizing the Hamiltonian with size $60\times 60$. }
\label{tableEnNH}
\end{table}

\begin{table}[h]
\begin{tabular}{|c|c c c|}
\hline
$\Delta\langle x\rangle_n^{(K)}$ & $K=1$ & $K=2$ & $K=3$  \\
\hline
$n=0$ &$-3\times10^{-2}i$& $2\times10^{-6}i$ & $-2\times10^{-11}i$  \\
$n=1$ &&$-8\times10^{-4}i$ &$1\times10^{-8}i$ \\
$n=2$ && &$2\times10^{-6}i$ \\
\hline
\end{tabular}
\caption{The null bootstrap predictions for $\langle x\rangle_n$ 
with the non-Hermitian potential $V(x)=ix^3$ and $L=K+1$. 
The errors are defined as $\D \langle x\rangle^{(K)}_n=\langle x\rangle^{(K)}_n-\langle x\rangle^\star_n$ 
and only the first digits are provided.  
The reference values $\langle x\rangle_0^\star=-0.590072533091i$, 
$\langle x\rangle_1^\star=-0.9820718380i$, 
$\langle x\rangle_2^\star=-1.20548075i$  
were obtained by diagonalizing the Hamiltonian of size $60\times 60$. }
\label{tablexNHx}
\end{table}

After solving the consistency relations \eqref{H-relation}, 
we can express $\<x^m p^n\>_E$ in terms of 
$\{E, \<x\>_E, \<1\>_E\}$. 
We again set $\<1\>_E=1$. 
However, this choice does not guarantee the non-negativity of the $\mathcal {PT}$ norm. 
As in the Hermitian example, the $\eta$ minimization for the non-Hermitian Hamiltonian \eqref{Hx3} 
also captures more excited states as $K$ increases. 
In Table \ref{tableEnNH} and \ref{tablexNHx}, 
we present the $\eta$ minimization results of the low-lying energy spectrum and the corresponding $\langle x\rangle=i\,\text{Im} \langle x\rangle$. 
\footnote{Since a $\mathcal P\mathcal T$ symmetric wavefunction satisfies
$[\psi(-x)]^\ast=\pm \psi(x)$, 
%so we have
%\be
%\text{Re}\, \psi(-x)-i\,\text{Im}\, \psi(-x)
%=
%\pm[\text{Re}\, \psi(x)+i\,\text{Im}\, \psi(x)]. 
%\ee
the real and imaginary parts of $\psi(x)$ have opposite parity.  
We also have $\langle 1\rangle_E=\pm C\int \mathrm{d}\, x\,[\psi_E(x)]^2=1$, 
$\langle x\rangle_E=\pm C\int \mathrm{d}x \, x\,[\psi_E(x)]^2$.
As a result, we expect that $C$ is a real number 
and $\langle x\rangle_E$ is purely imaginary for $\mathcal{PT}$ symmetric solutions. 
}
Again, the precision increases rapidly with the cutoff parameter $K$. 
We set $a_{00}=0$  at $K=3$ to avoid the trivial solution $\mathcal O^{\text{trivial}}_\text{null}=H-E$.

Above, we successfully derived the real spectrum of 
the complex Hamiltonian \eqref{Hx3} using the $\eta$ minimization. 
A curious question is: Are there $\eta$ minima with complex energies?  
It turns out that there indeed exist conjugate pairs of local minima with complex $E$ and complex $\langle x\rangle_E$. 
\footnote{This is also related to the fact that we do not introduce the complex conjugate of $E$ in \eqref{H-relation}. 
A more systematic study should treat the imaginary part of every variable more carefully. }
These minima persist as we increase $L-K$, so they should not be numerical coincidences. 
They might be related to other boundary conditions for the wave functions, 
but we do not fully understand the physical implications of these complex solutions. 

\section{Ladder operators and matrix elements}
\label{sec-ladder-ME}
In the previous section, we have derived the precise values of low-lying spectra $E_n$ and 
expectation values $\langle\psi_E|x^mp^n|\psi_E\rangle$ using the null bootstrap. 
Can we also compute the non-diagonal matrix elements? 
\footnote{This question was inspired by the interesting work \cite{Nancarrow:2022wdr}. } 
To address this question, 
let us remind the reader that we have introduced the ladder operators in Sec. \ref{sec-null-bootstrap-QM}, 
which generate different energy eigenstates from a given one
\be
|\psi_{E^\prime}\rangle=L_{E^\prime E}|\psi_{E}\rangle\,.
\ee
If one can determine the expressions of ladder operators, 
then the matrix elements are not independent parameters
\be
\langle\psi_E|x^mp^n|\psi_{E^\prime}\rangle
=
\langle x^mp^n L_{E^\prime E}\rangle_E\,.
\ee
We will again approximate the ladder operators by finite degree polynomials
\be\label{ladder-polynomial}
L^{(M)}_{E^\prime E}=
\sum_{m=0}^{M}\sum_{n=0}^{M-m} c_{m n }\,x^{m}(ip)^{n}\,.
\ee
The coefficients $c_{mn}$ are determined by the $\eta$ minimization for the null constraints
\footnote{The inner product $\langle \psi_\text{test}^{(L)}|\psi_\text{null}^{(K)}\rangle$ in \eqref{eta-fun} is replaced by $\langle \psi_\text{test}^{(L)}|(H-E^\prime)L^{(M)}_{E^\prime E}|\psi_E\rangle$.}
\be
\langle \psi_\text{test}^{(L)}|(H-E^\prime)L^{(M)}_{E^\prime E}|\psi_E\rangle\approx 0\,,
\ee
which is an approximate version of the Schr\"{o}dinger-like equation \eqref{Schrodinger}. 
We will choose a normalization such that the absolute norm of $|\psi_{E^\prime}\rangle$ is one and 
the signs of the lowest non-diagonal matrix elements are positive.

In Fig.\ref{tableME}, we present some results of the quartic potential \eqref{H4} 
based on the $K=4$ results. %  in Sec. \ref{application}. 
Naturally, the low-lying observables are more precise than the high-lying ones. 
For $M<K$, one can see that the results improve with the approximation order $M$. 
If $M> K$, the precision will decrease 
as the input parameters were derived from a truncation of order-$K$.  

\begin{table}[h]
\begin{tabular}{|c|c c c c|}
\hline
$\Delta \langle 0 | x^m |n\rangle^{(M)}$ &  $m=1$ &$m=2$ &  $m=3$& $m=4$  \\
\hline
$n=1, M=1$  &$4\times 10^{-6}$& &$1\times 10^{-5}$& \\
$n=2, M=2$ &&$3\times 10^{-7}$& &$1\times 10^{-6}$ \\
$n=1, M=3$ &$1\times 10^{-11}$& &$2\times 10^{-11}$& \\
\hline
\end{tabular}
\caption{The null bootstrap predictions 
for the low-lying matrix elements of anharmonic oscillator with the Hermitian potential $V(x)=x^2+x^4$. 
We use a shorthand notation $\langle 0 | x^m|n\rangle=\langle \psi_{E_0} | x^m|\psi_{E_n}\rangle$. 
The differences are defined as $\Delta \langle 0 | x^m |n\rangle^{(M)}=\langle 0 | x^m |n\rangle^{(M)}-\langle 0 | x^m |n\rangle^{\star}$. 
The reference values from the Hamiltonian diagonalization are 
$\langle 0 | x^1 |1\rangle^\star=0.552565959314$, 
$\langle 0 | x^3 |1\rangle^\star=0.456180404562$, 
$\langle 0 | x^2 |2\rangle^\star=0.406699817396$, 
$\langle 0 | x^4 |2\rangle^\star=0.62270464561$. 
Some matrix elements should vanish due to the unbroken parity symmetry of the energy eigenstates. 
Their numerical values are of order $10^{-11}$ or smaller, which are not presented here. 
}
\label{tableME}
\end{table}

Since the parity symmetry of \eqref{H4} is unbroken, 
the ladder operators have numerically small coefficients for the parity violating terms. 
For example, the numerical coefficients of $1, x^2, p^2, xp$ in $L_{E_1,E_0}^{(M=3)}$ are of order $10^{-15}$. 
We do not present the results of $n=1, M=2$ in Table \ref{tableME}, because the precision is similar to the case of $n=1, M=1$. 
These parity constraints also explain the precision pattern of Table \ref{tableEn} and \ref{tablex2} as we increase $K$. 
For consistency, the parity-forbidden matrix elements indeed take negligible numerical values. 
In addition, we have also verified the approximate orthogonality of different eigenstates, 
i.e. $\langle 0|n\rangle^{(M)}\approx 0$ for $n>0$.

For the completeness of eigenfunctions, it is natural that the party of $|n\rangle$ is given by 
$(-1)^n$ as in the harmonic case. 
Then the level-$j$ ladder operator that connects $|n\rangle$ and $|n+j\rangle$ 
should consist of only parity-even (odd) monomials for even (odd) $j$. 
This also applies to the non-Hermitian case of $V(x)=i g x^3$:   
The terms violating $\mathcal {PT}$-symmetry make negligible contributions to the ladder operators, 
as they should be composed of monomials with the same $\mathcal {PT}$ eigenvalues, 
up to a phase factor
\footnote{One should be careful about the phase as $i\rightarrow -i$ 
under the action of the $\mathcal T$ operator.  
The $\mathcal {PT}$ norm is not sensitive to the phase choice. }.
By taking into account the global symmetry, 
the polynomial ansatz of \eqref{ladder-polynomial} takes a more restricted form.

We can further study the algebraic properties of the ladder operators using their approximate expressions. 
Although $L^{(M)}_{E E^\prime}L^{(M)}_{E^\prime E}$ are finite degree polynomials with non-zero expectation values, 
we observe that
\be
\langle\mathcal O^{(L)}_\text{test}\,[H, L^{(M)}_{E E^\prime}L^{(M)}_{E^\prime E}]\rangle_E\approx0\,,
\ee
so these commutators are close to a linear combination of null operators. 
Note that $L^{(M)}_{E E^\prime}$ is not derived from the Hermitian conjugation of $L^{(M)}_{E^\prime E}$. 
The action of $L_{E^\prime E}$ and then $L_{E E^\prime}$ on $|\psi_E\rangle$ should leave it invariant up to some numerical factor.  
These product operators $L_{E E^\prime}L_{E^\prime E}$ can be interpreted 
as the interacting counterparts of the number operator of harmonic oscillator.

\section{Discussion}
In this work, we proposed the principle of nullness for bootstrapping quantum systems. 
The existence of many null state conditions can guarantee that the energy spectrum is bounded from below, 
so the quantum system is stable. 
\footnote{One may further impose stability under deformations of the Hamiltonians. }
As concrete examples, 
we use the $\eta$ minimization to obtain high precision results for 
both Hermitian and non-Hermitian anharmonic oscillators. 

It would be interesting to apply the null bootstrap to quantum many body problems. 
Recently, the positive bootstrap has made some initial steps in this direction \cite{Han:2020,Lawrence:2021msm}. 
Since we argue that the null bootstrap can deduce the positive bootstrap results, 
it should also furnish a new method for studying many important but challenging lattice models, 
ranging from high energy physics to condensed matter physics. 

Previously, we constrained the Ising and Yang-Lee CFT data by imposing the operator decoupling 
in the intermediate spectrum of identical correlators \cite{Li:2017agi}.   
In light of the results in this work, 
it would be interesting to consider a system of correlators 
such that their superposition gives rise to an external null state. 
This approach can be viewed as a $d>2$ generalization of the Virasoro null state constraints on correlators  
that lead to the Belavin-Polyakov-Zamolodchikov differential equations. 

In the end, we list some questions that we find  interesting:
\begin{itemize}
\item
Can we interpret the large number of null state conditions as a kind of ``integrability''? 
\footnote{In \cite{Dorey:1998pt}, a connection was proposed between ordinary differential equations and integrable models, 
based on the similarity of functional relations in the two contexts.  }
It would be interesting to further study the algebraic properties of the null and ladder operators. 
\footnote{In 2d CFTs, they are related to the chiral algebra, such as the Virasoro symmetry. 
For free field theories, they are also associated with the higher spin symmetry. 
There may exist a nonlocal, higher generalization, 
closely related to the Regge continuation. 
See \cite{Binder:2019zqc} for the categorical interpretation of $O(N)$ symmetry with non-integer $N$.  }
At the end of Sec. \ref{sec-ladder-ME}, we presented some preliminary results concerning this question. 
\item
Our bootstrap study is based on the operator-algebra formulation 
and we do not use the explicit expressions of the eigenfunctions. 
What are the algebraic counterparts of many important concepts in the wave function formulation, 
such as boundary conditions, Stokes sectors and WKB method?
\footnote{See \cite{Nakayama:2022ahr} for a discussion about the classical limit $\hbar\rightarrow 0$ in the bootstrap context. }
\item
The non-Hermitian models can be extended to non-integer powers. 
Can we also study them from the bootstrap perspective? 
This will require the introduction of non-integer ``power'' to the operator algebra. 
This is closely related to the analyticity of the expectation values. 
\item
Can a physical operator algebra has a bounded-from-below representation without resorting to some null state conditions? 
\footnote{A candidate Hamiltonian is $H=(a_-a_+)^2$ with $[a_-,a_+]=\hbar$. }
So the vacuum state is NOT annihilated by any lowering operators? 

\item
Projection techniques play a significant role in both high energy physics and condensed matter physics, 
such as the GSO projection in superstring theories \cite{Gliozzi:1976qd} and 
projective construction of quantum spin-liquid states \cite{Wen:2004ym}. 
Can we view them as special cases of the nullness principle and make more general classifications? 
\footnote{For 2d minimal-model CFTs,  the various global symmetries are consequences of the null classification. } 
\end{itemize}

\begin{acknowledgments}
I am grateful to Shuai Yin for discussions. 
I also thank Sachin Jain and Sakil Khan for correspondences and the referees for the constructive comments. 
This work was partly supported by the 100 Talents Program of Sun Yat-sen University 
and the Fundamental Research Funds for Central Universities, Sun Yat-sen University (22qntd3005). 
\end{acknowledgments}

\end{document}